\documentclass[preprint,twocolumn,showpacs,preprintnumbers,amsmath,amssymb,superscriptaddress,10pt]{revtex4}
\usepackage{latexsym}
\usepackage{graphicx}
\usepackage{dcolumn}
\usepackage{mathrsfs}

\usepackage{color}

\newcommand{\green}[1]{}
\newcommand{\blue}[1]{}

\begin{document}
\title{Evolution of cooperation is a robust outcome
in the prisoner's dilemma on dynamic networks}
\author{Yoshimi Yoshino}
\email{yoshimi@stat.t.u-tokyo.ac.jp}
\affiliation{Department of Mathematical Informatics,
  the University of Tokyo, 7-3-1 Hongo, Tokyo 113-8656, Japan}
\author{Naoki Masuda}
\email{masuda@mist.i.u-tokyo.ac.jp}
\affiliation{Department of Mathematical Informatics,
  the University of Tokyo, 7-3-1 Hongo, Tokyo 113-8656, Japan}
\affiliation{PRESTO, Japan Science and Technology Agency, 
4-1-8 Honcho, Kawaguchi, Saitama 332-0012, Japan}
%
%
%
%
\begin{abstract}
Dynamics of evolutionary games strongly depend on underlying
networks. We study
the coevolutionary prisoner's dilemma in which
players change their local networks as well as strategies
(i.e., cooperate or defect). This topic has been
increasingly explored by many researchers.  On the basis of active linking
dynamics [J. M. Pacheco et al., J. Theor. Biol. 243, 
437 (2006), J. M. Pacheco et al., Phys. Rev. Lett. 97, 
258103 (2006)], we show that 
cooperation is enhanced fairly robustly. In particular, cooperation evolves
when the payoff of the player
is normalized by the number of neighbors; 
this is not the case in the evolutionary prisoner's dilemma on
static networks.
\end{abstract}
%
%
\maketitle

\section{Introduction}\label{sec:intro}

Existing studies have mainly focused on the mechanisms underlying
other-regarding behavior in
social dilemma situations.
A prototypical model for studying this subject is
the prisoner's dilemma game, where each player either cooperates 
or defects. From an egoistic perspective, defection is more
lucrative and is the unique Nash equilibrium.
However, mutual cooperation is more profitable for a population.
Maintenance and emergence of cooperation in the prisoner's dilemma
are often explained
in terms of evolutionary dynamics, where individuals imitate
successful others in the social dynamics nomenclature
\cite{Axelrodbook,Nowak06book}.

Upshots of evolutionary game dynamics generally
depend on the network structure underlying interaction between players.
In particular, it was recently discovered that
heterogeneous networks enable
cooperation in the evolutionary
prisoner's dilemma
\cite{SantosPacheco05prl,SantosPacheco2006,Santosetal06pnas,Duran_Mulet_2005}.
In this scheme,
a hub (i.e., a player directly connected to many others)
tends to earn higher payoffs than a player with a small degree (i.e.,
the number of neighbors for a player).
Cooperation on hubs can be 
stabilized, whereas defection on hubs cannot.
Cooperation then propagates from hubs to the periphery players.
A less-noticed fact underlying these results is 
that the payoff for each player is defined as the sum of the payoffs
earned by playing against all the neighbors.
If we divide the payoff by the player's degree
\cite{SantosPacheco2006,Tomassinietal2007,Szolnoki20082075},
which we call the average payoff,
or if we shift each element of the payoff matrix by a constant
\cite{Masuda07prsb,Luthi2008955}, cooperation is not enhanced
in heterogeneous networks. Therefore, for these definitions,
hubs are not advantageous in terms of the payoff itself.

It may be realistic to consider that players dynamically change partners.
Indeed, when strategies of players and the network structure
coevolve such that players preferentially
link to cooperators, cooperation on a global scale is facilitated
\cite{Zimmermann04,Zimmermann05,EguiluzZCC05ajs,Pacheco_etal_2006a,Pacheco_etal_2006b,SantosPL06plos}
%
%
(also see 
\cite{Szabo200797,GrossBlasius08,PercSzolnoki10biosys} for reviews).
Most previous studies on coevolutionary social dilemma
dynamics on networks focused on the summed payoff scheme, in which
well-connected players tend to obtain a large payoff.
In contrast, the present study examines the coevolutionary prisoner's dilemma
in a more adverse condition for cooperators: the
average payoff scheme. We show that
even under the average payoff scheme, cooperation emerges through
coevolutionary dynamics, with the exception of a specific update rule
that is known to disfavor cooperation in evolutionary games
on static networks.

\section{Model}\label{sec:model}
\subsection{Prisoner's dilemma dynamics}

We consider the prisoner's dilemma game in which each player at a node
is either a cooperator (C) or a defector (D) in each round.  We use
$s_i\in \{ {\rm C}, {\rm D} \}$ to denote the strategy selected by the $i$th player.

We set the payoff matrix to \cite{Pacheco_etal_2006b}
\begin{equation}
\bordermatrix{ 
  & \mbox{C} & \mbox{D} \cr 
\mbox{C} & 0.5 & -0.5 \cr 
\mbox{D} & 1 & 0 } 
\end{equation}
The values represent the payoff that the row player obtains.
We assume the symmetric prisoner's dilemma game; the payoff of
the column player is determined analogously.
For example, if player $i$ cooperates and player $j$ defects,
$i$ and $j$ gain $-0.5$ and 1, respectively.

In one round, each player plays the prisoner's dilemma against
all the neighbors. We consider two types of 
collective payoff. The first is the
summed payoff $P_i$ defined for the $i$th player
as the summation of the payoff
gained over all the neighbors. The second is the
average payoff $P_i/(k_i/\left<k\right>)$,
where $k_i$ is the degree of node $i$ and 
$\left<k\right>$ is the mean degree of the network. 
The normalization factor
$\left<k\right>$ is used to make the magnitudes of the summed and
average payoffs similar.
On static heterogeneous networks, the summed payoff
scheme enhances cooperation \cite{SantosPacheco05prl,SantosPacheco2006,Santosetal06pnas,Duran_Mulet_2005}, whereas
the average payoff scheme does not
\cite{Luthietal05,Tomassinietal06}.

We assume that the players alter the strategy
according to the so-called
Fermi update rule \cite{SzaboToke98,Traulsen06pre,SantosPL06plos},
unless otherwise
stated. According to Fermi rule, each player
is selected at the rate $1/T_{\rm s}$. 
The selected player, denoted as $i$,
chooses one of his/her neighbors, denoted as $j$.
The strategy of
the $j$th player replaces that of the $i$th player with probability
$1/(1+\exp[\beta(P_i-P_j)])$. Otherwise, the strategy of the
$i$th player replaces that of the $j$th player.
We set $\beta=1$.
In the summed payoff scheme,
the fraction of cooperators is large for a small
$\beta$ \cite{Pacheco_etal_2006b}. The main focus 
of the following numerical simulations is
the average payoff scheme.

\subsection{Dynamics of networks}

The network is assumed to evolve in accordance with
the active linking (AL) model proposed by Pacheco and colleagues
\cite{Pacheco_etal_2006a,Pacheco_etal_2006b}. We implement the AL model
as follows:
\begin{enumerate}
\item Each pair of players $i$ and $j$ is selected at the
rate $1/T_{\rm a}$. 

\item  If players $i$ and $j$ are not adjacent,
we connect them with probability $\alpha_{s_i}\alpha_{s_j}$. 
If nodes $i$ and $j$ are adjacent,
we disconnect them with probability $\gamma_{s_{i}s_{j}}$. 
If $k_i=1$ or $k_j=1$, we do not disconnect them to keep the network
connected.
\end{enumerate}
We set 
$\alpha_{\mbox{\scriptsize C}} =\alpha_{\mbox{\scriptsize D}}=0.15, \gamma_{\mbox{\scriptsize CC}}=0.1, \gamma_{\mbox{\scriptsize CD}}=\gamma_{\mbox{\scriptsize DC}}=0.8$, and $\gamma_{\mbox{\scriptsize DD}}=0.32$, unless otherwise stated.
The values of $\alpha_{\rm C} $ and $\alpha_{\rm D}$ are smaller than those
used in \cite{Pacheco_etal_2006a,Pacheco_etal_2006b}. As a result,
relatively sparse networks emerge.
This is aimed at emphasizing
heterogeneity in the degree and examining its differential effects in
the summed and average payoff schemes.

At the meanfield level, the AL dynamics are described by 
\begin{equation}
\dot{X}_{s_i s_j} = \frac{1}{T_{\rm a}} \{ \alpha_{s_i}  \alpha_{s_j} (X_{s_i s_j}^{
  \mbox{\scriptsize max}}-X_{s_i
  s_j}) - \gamma_{s_i s_j} X_{s_i s_j} \}, 
\end{equation}
where $(s_i, s_j)=(\mbox{C}, \mbox{C}), (\mbox{C} , \mbox{D})$, or $(\mbox{D}, \mbox{D})$, $X_{s_i s_j}$ is the number
of links that have $s_i$ and $s_j$ at the two ends, and
$X_{s_i s_j}^{\max}$
$=N(N-1)/2$.

\subsection{Setup for numerical simulations}

We assume $N=100$ players. 
The players initially form the complete graph, unless otherwise
stated.
The strategy update and the AL dynamics are defined as independent
Poisson processes. Because we focus on the stationary state of the
coevolutionary dynamics, we set $T_{\rm s}=1$
without loss of generality
\cite{Pacheco_etal_2006b}.
We stop each realization at $T^\infty = 200\times \max\{T_{\rm a},
T_{\rm s}\}$ and regard the final state as an approximate stationary state. 
The quantities shown in the following sections
are the averages over $10$ realizations.


\section{Results}\label{sec:results}

\subsection{Evolution of cooperation for different payoff schemes}

\begin{figure}[htbp]
\begin{center} 
  \begin{tabular}{cc}

\vspace{1cm}

\hspace{-2.cm}
     \rotatebox{-90}{\resizebox{60mm}{!}{\includegraphics{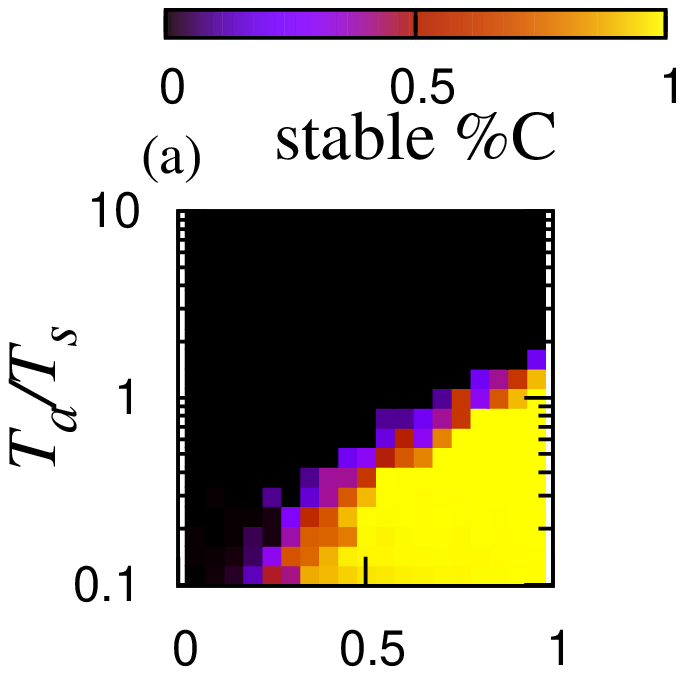}}}
\hspace{-4.5cm}
     \rotatebox{-90}{\resizebox{60mm}{!}{\includegraphics{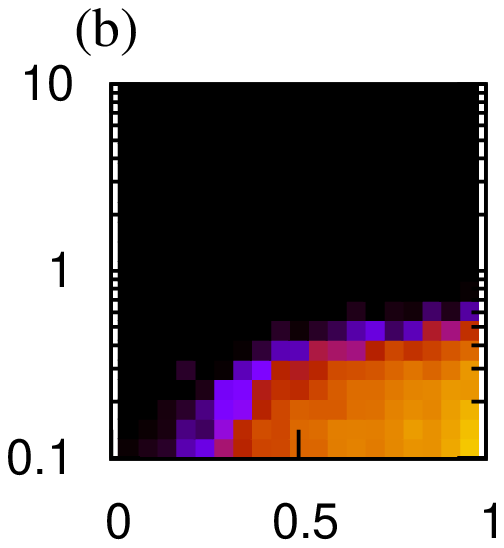}}}\\

\vspace{1cm}

\hspace{-2cm}
     \rotatebox{-90}{\resizebox{60mm}{!}{\includegraphics{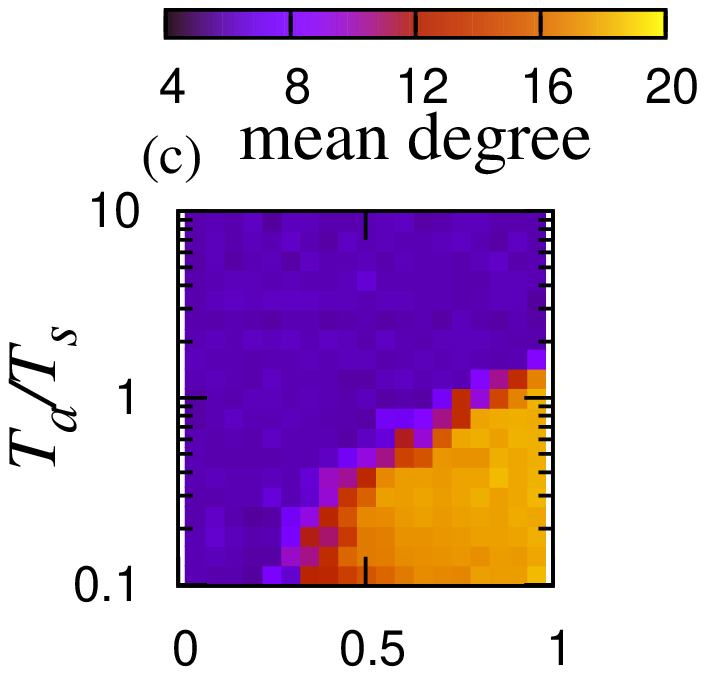}}}
\hspace{-4.5cm}
     \rotatebox{-90}{\resizebox{60mm}{!}{\includegraphics{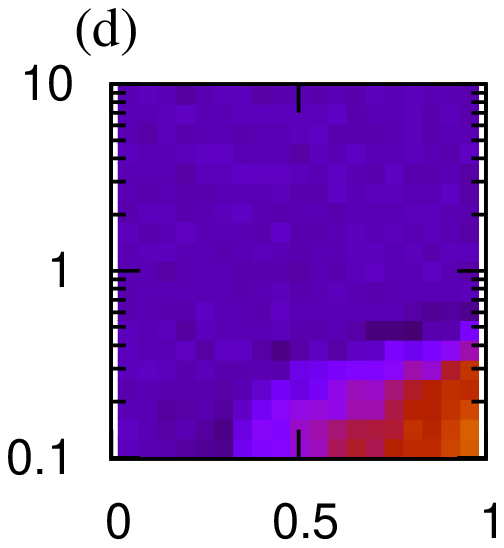}}} \\
\hspace{-2cm}
     \rotatebox{-90}{\resizebox{60mm}{!}{\includegraphics{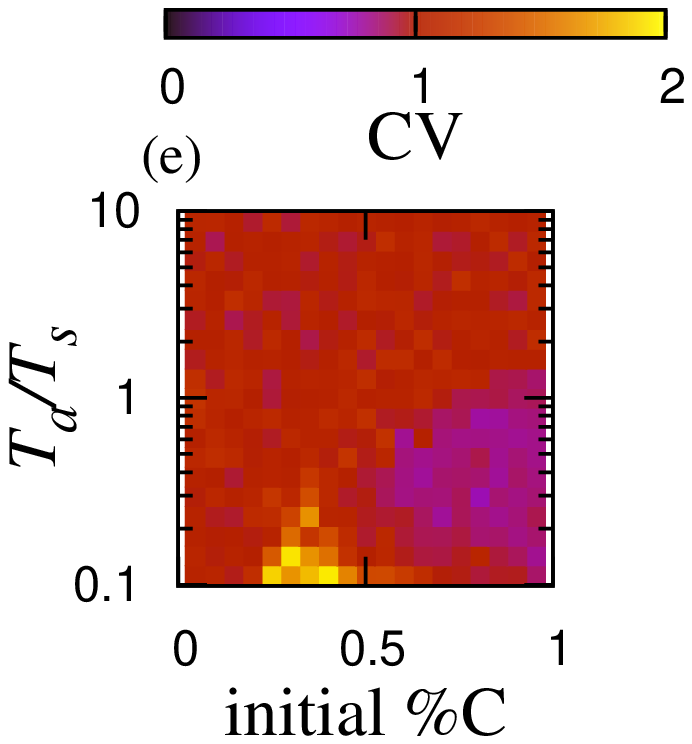}}}
\hspace{-4.5cm}
     \rotatebox{-90}{\resizebox{60mm}{!}{\includegraphics{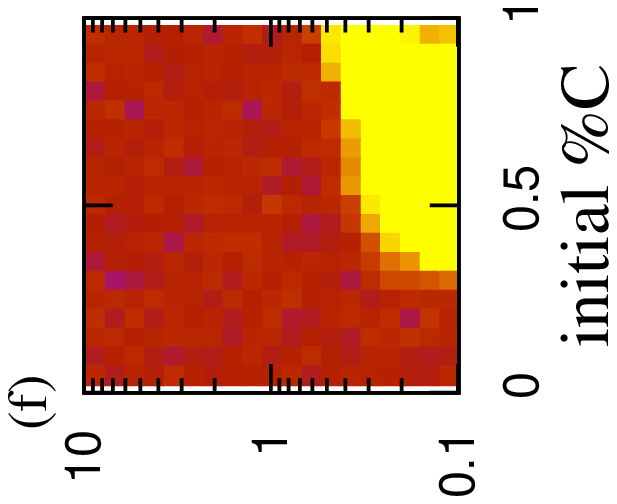}}} \\
  \end{tabular}

\caption{(Color online) Results for 
 the ordinary AL dynamics with Fermi update rule in the final state.
(a, c, e) Summed payoff scheme.
(b, d, f) Average payoff scheme.
(a), (b) Fraction of cooperators. (c), (d)
Mean degree. (e), (f)
CV of the degree distribution.
}\label{fig:al}
\end{center}
\end{figure}

The final fraction of cooperators
is shown in Fig.~\ref{fig:al}(a) and \ref{fig:al}(b) 
for the summed and average payoff schemes, respectively.
The results for the summed payoff scheme (Fig.~\ref{fig:al}(a))
are consistent with those in \cite{Pacheco_etal_2006a,Pacheco_etal_2006b};
cooperators flourish if the AL is fast enough relative to the strategy update
and many of them exist initially.
We find that cooperation also survives in the average
payoff scheme (Fig.~\ref{fig:al}(b)),
although it is slightly more difficult to maintain
than in the case of the summed payoff scheme.

Because the two payoff schemes coincide when the network is regular
(i.e., homogeneous in the degree), we examine
the dispersion in the degree.
The mean degree of the nodes in the final networks
is shown in Fig.~\ref{fig:al}(c) and
\ref{fig:al}(d) for the summed (Fig.~\ref{fig:al}(a))
and average (Fig.~\ref{fig:al}(b)) payoff schemes, respectively.
The coefficient of variance (CV)
of the degree, defined by $\sum_{i=1}^N \left(k_i-\left<k \right>\right)^2/(\left<k\right>N)$, in the final networks
is shown for the summed and average payoff schemes in
Figs.~\ref{fig:al}(e) and \ref{fig:al}(f), respectively.
We obtain CV $\approx$ 1 for both payoff schemes, suggesting that
the dispersion of the degree is 
comparable to that implied by the Poisson distribution.

Our main finding in this section is that cooperation is maintained
regardless of the payoff scheme if coevolutionary dynamics are considered.
We examine the robustness of this finding in the following.

\subsection{Heterogeneous networks}

Networks with heterogeneous degree distributions
do not enhance cooperation if they are static and 
the average payoff scheme is used
\cite{SantosPacheco2006,Tomassinietal2007,Szolnoki20082075}.
Although coevolutionary dynamics with the average payoff
scheme enhance cooperation, the Poisson degree distribution
revealed in Fig.~\ref{fig:al} 
may not be heterogeneous enough to sufficiently distinguish between
the consequence of the summed payoff scheme and that of the average
payoff scheme.
Therefore, we perform
additional numerical simulations with a modified AL model that yields
more heterogeneous networks.

We use a variation of the static network model proposed by Goh and
colleagues \cite{Gohetal2001}. In the original network model,
nodes $i$ and $j$ ($1\le i, j\le N$) are connected with probability
proportional to $w_i w_j$, where $w_i = i^{-\alpha}$.  The obtained
network has degree distribution $p(k) \propto k^{-\gamma}$, where
$\gamma=1+1/\alpha$. On the basis of this model, we modify the AL
dynamics by replacing the rate at which a link is created between
players $i$ and $j$, i.e. $\alpha_{s_i} \alpha_{s_j}$ by $w_i w_j
\alpha_{s_i} \alpha_{s_j} / \mathscr{N}$. We set the normalization constant
$\mathscr{N}=0.0015$ so that the average degree in the final
network is comparable to that for the original AL dynamics.
The rule for removing links remains unchanged.

\begin{figure}[!t]
\begin{center} 
  \begin{tabular}{cc}
\vspace{1cm}

\hspace{-2cm}
     \rotatebox{-90}{\resizebox{60mm}{!}{\includegraphics{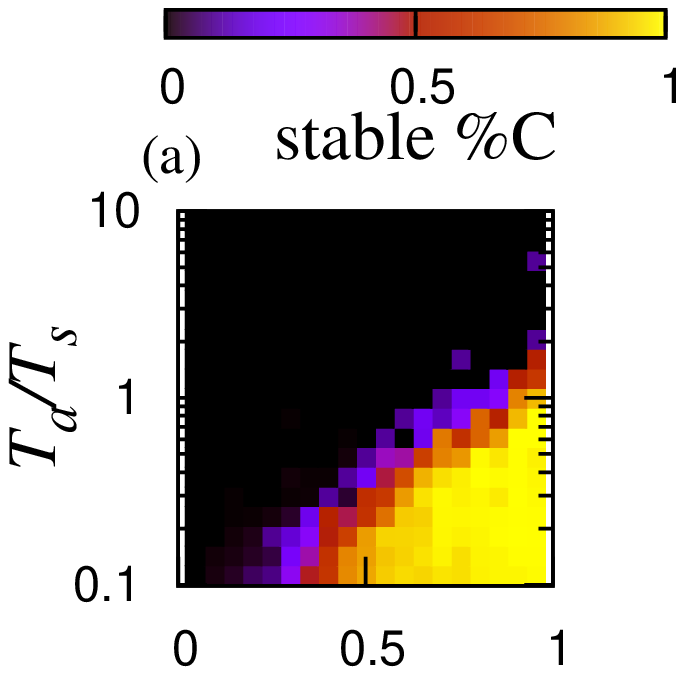}}}
\hspace{-4.5cm}
     \rotatebox{-90}{\resizebox{60mm}{!}{\includegraphics{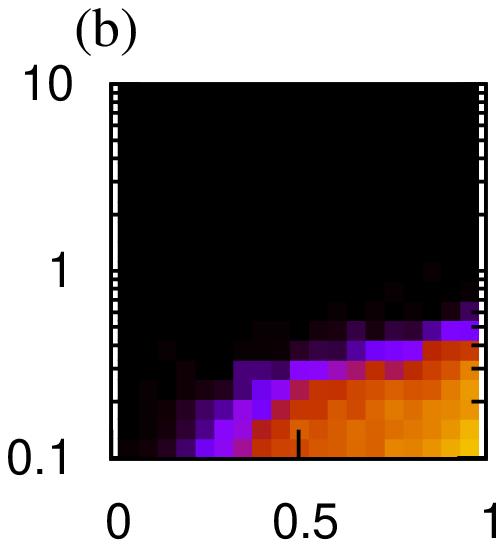}}}\\

\vspace{1cm}

\hspace{-2cm}
     \rotatebox{-90}{\resizebox{60mm}{!}{\includegraphics{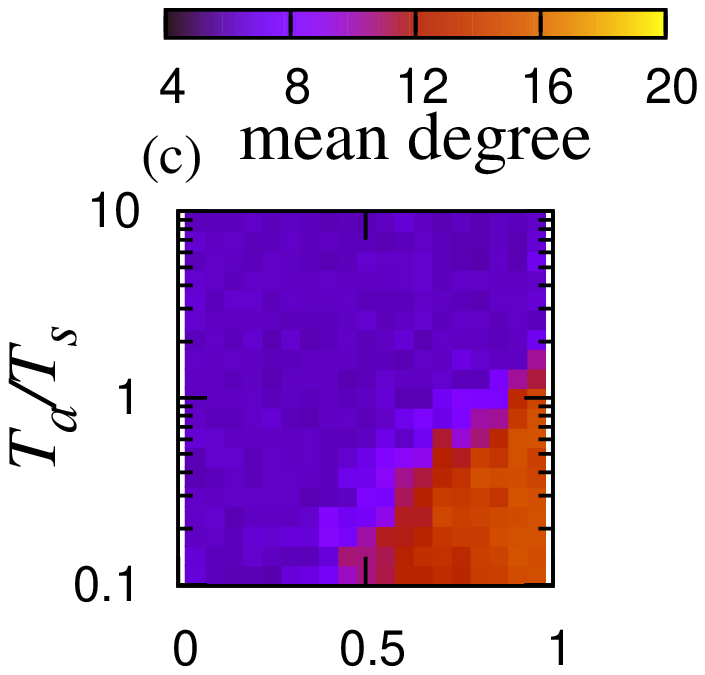}}}
\hspace{-4.5cm}
     \rotatebox{-90}{\resizebox{60mm}{!}{\includegraphics{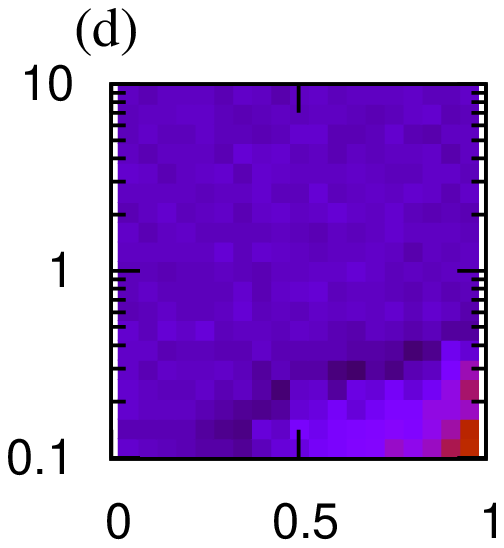}}} \\
\hspace{-2cm}
     \rotatebox{-90}{\resizebox{60mm}{!}{\includegraphics{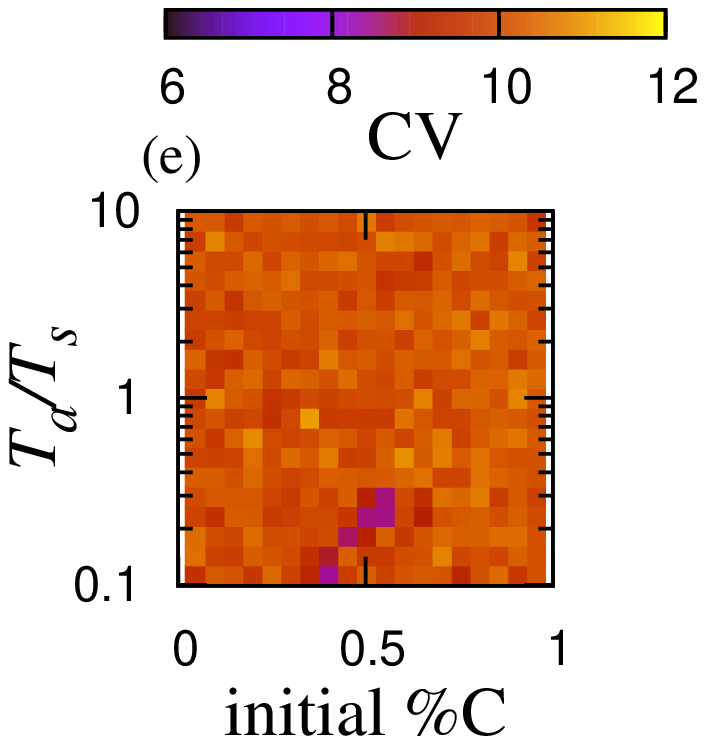}}}
\hspace{-4.5cm}
     \rotatebox{-90}{\resizebox{60mm}{!}{\includegraphics{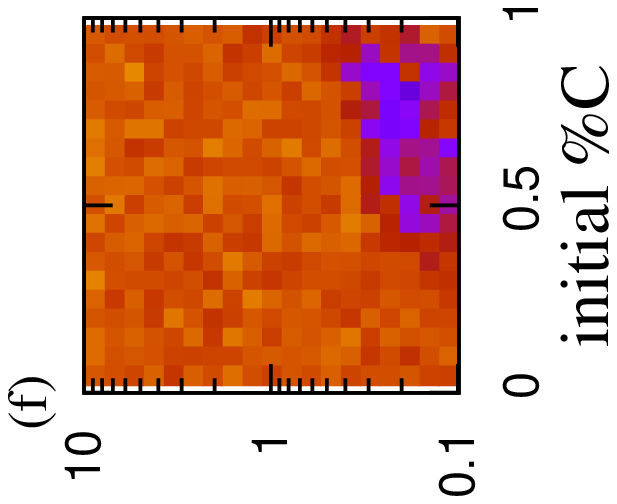}}} \\
  \end{tabular}

\caption{(Color online) Results for the 
modified AL dynamics. See the caption of
Fig.~\ref{fig:al} for the legends.}\label{fig:goh}
\end{center}
\end{figure}

The numerical results for this variant of the AL model are 
shown in Fig.~\ref{fig:goh}. The fraction of
cooperation is approximately the same as that of the original
AL model (Fig.~\ref{fig:al}(a), (b)); cooperation emerges in both the summed
(Fig.~\ref{fig:goh}(a)) and average (Fig.~\ref{fig:goh}(b)) 
payoff schemes
when AL is fast and sufficient cooperators exist initially.
As planned, the average degree in the final networks
(Fig.~\ref{fig:al}(c), (d)) is comparable to
that obtained from the original AL model.
As shown in Fig. \ref{fig:goh}(e, f),
the degree distribution is considerably more heterogeneous 
than in the case of the original AL model
(Fig. \ref{fig:al}(e, f)).
Coevolutionary dynamics in the average payoff scheme
enhance cooperation for both homogeneous and heterogeneous networks.

\subsection{Different update rules}\label{sub:different update rules}
\begin{figure}[!t]
\begin{center} 
  \begin{tabular}{cc}

\vspace{-1.4cm}

\hspace{-2cm}
     \rotatebox{-90}{\resizebox{60mm}{!}{\includegraphics{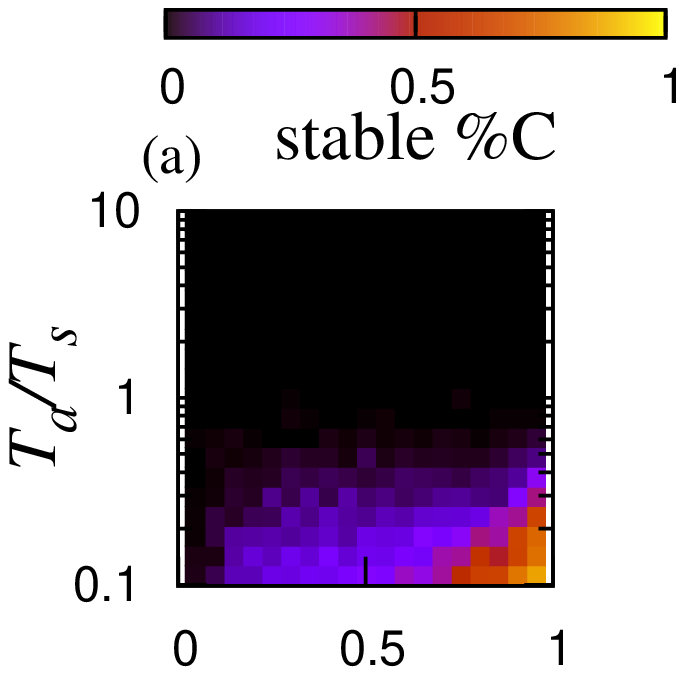}}}
\hspace{-4.5cm}
     \rotatebox{-90}{\resizebox{60mm}{!}{\includegraphics{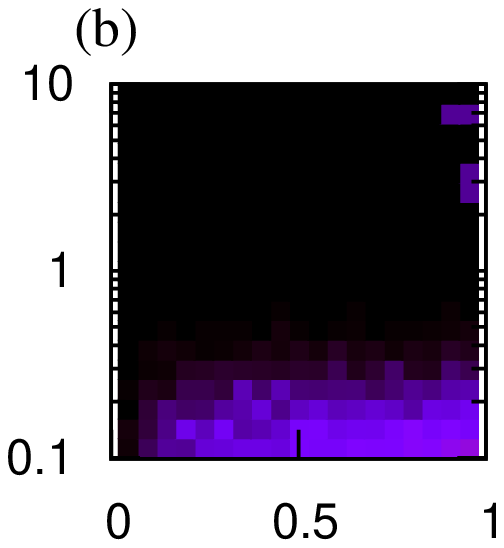}}}\\

\vspace{-1.4cm}

\hspace{-2cm}
     \rotatebox{-90}{\resizebox{60mm}{!}{\includegraphics{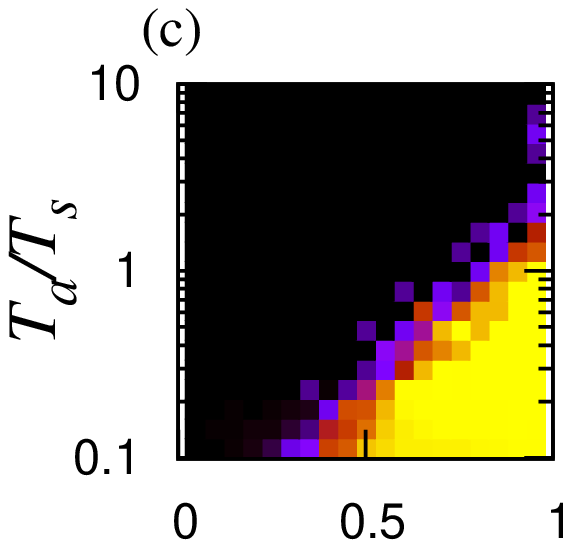}}}
\hspace{-4.5cm}
     \rotatebox{-90}{\resizebox{60mm}{!}{\includegraphics{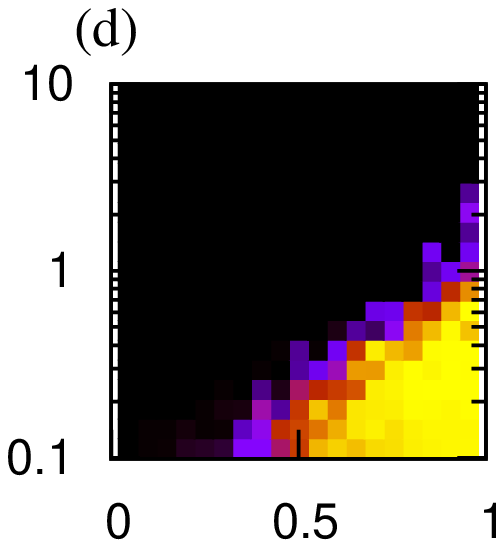}}} \\
\hspace{-2cm}
     \rotatebox{-90}{\resizebox{60mm}{!}{\includegraphics{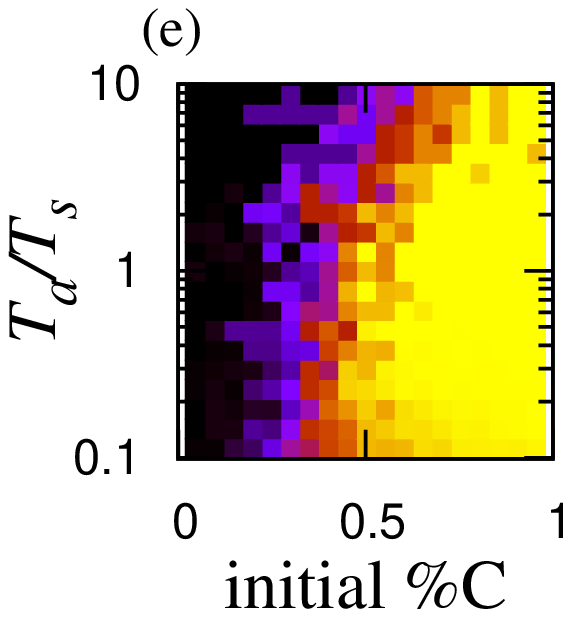}}}
\hspace{-4.5cm}
     \rotatebox{-90}{\resizebox{60mm}{!}{\includegraphics{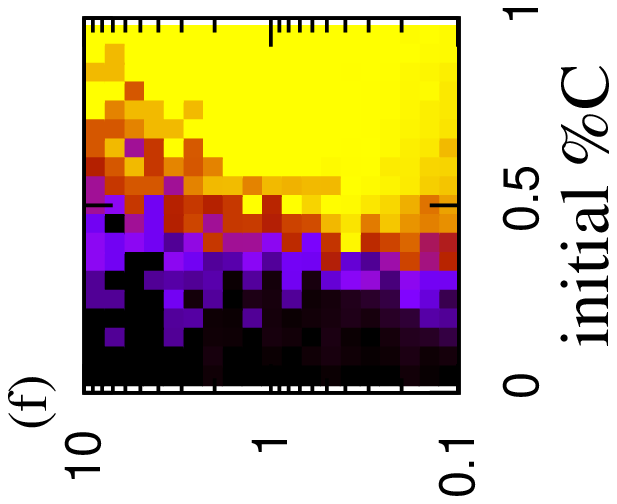}}} \\
  \end{tabular}

\caption{(Color online) Final fraction of cooperators for various
update rules. We use (a, c, e) summed and (b, d, f) average
payoff schemes. We use BD rule in (a) and (b),
DB rule in (c) and (d), and Nowak's rule with
$\delta=20$ in (e) and (f).
}\label{fig:update}
\end{center}
\end{figure}

The difference in update rules can drastically
affect evolutionary dynamics on static networks
\cite{Ohtsuki06nature,Antal06prl,OhtsukiNowak08jtb,SoodAR08pre,Masuda2009323}.
In this section, we examine the effects of different update rules
on coevolutionary dynamics.

We implement birth-death (BD) update rule
\cite{Nakamaru97,Ohtsuki06nature,Ohtsuki200686},
death-birth (DB) update rule \cite{Nakamaru1998101,Ohtsuki200686},
and a variant of the DB rule proposed by Nowak \cite{NowakBM94pnas}.
In BD rule, a player is selected for reproduction
with probability proportional to the
payoff, and his/her strategy is transmitted to a randomly chosen neighbor.
In DB rule, a player, selected with equal probability $1/N$ dies,
and the neighbors compete for reproduction on this node, such that
the reproduction probability is proportional to the payoff.
In Nowak's rule,
the reproduction probability of the neighbors in
the DB rule is assumed to be proportional to $P_i^{\delta}$ and
$\left(P_i/(k_i/\left<k\right>)\right)^{\delta}$ for the summed and
average payoff schemes, respectively.
The case $\delta=1$ is equivalent to
DB rule. We set $\delta=20$.

The final fractions of cooperators for the BD rule
are shown in Fig.~\ref{fig:update}(a) and 
\ref{fig:update}(b) for the summed and average payoff schemes,
respectively. In this case, it is difficult to achive cooperation.
This is consistent with the fact that cooperation is generally
not likely for BD rule in static networks, as compared to
other update rules \cite{Ohtsuki06nature,Ohtsuki200686}.
In contrast,
in DB rule (Fig. ~\ref{fig:update} (c, d)) and
Nowak's rule
(Fig. ~\ref{fig:update} (e, f)),
cooperation emerges when AL is fast and 
sufficient cooperators exist initially.

\subsection{Fragile C-C links}\label{sub:cgeqd}

\begin{figure}[!t]
\begin{center}
  \begin{tabular}{cc}
\hspace{-2cm}
     \rotatebox{-90}{\resizebox{60mm}{!}{\includegraphics{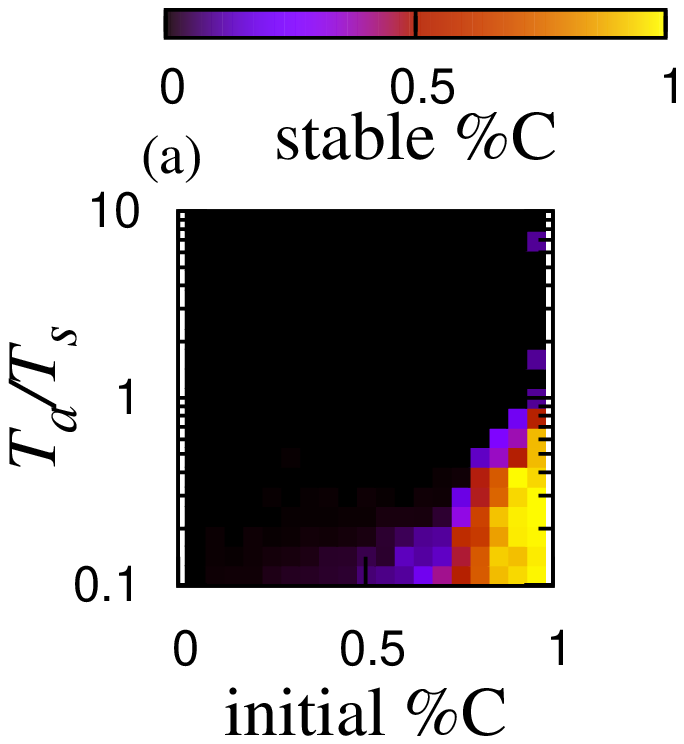}}}
\hspace{-4.5cm}
     \rotatebox{-90}{\resizebox{60mm}{!}{\includegraphics{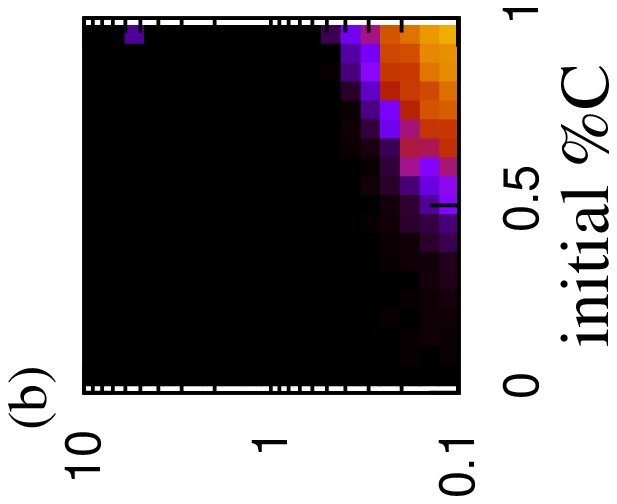}}} \\
  \end{tabular}
\caption{(Color online) Final fraction of cooperators when
$\gamma_{CC}=0.32$ and $\gamma_{DD}=0.1$.
We use Fermi update rule.
(a) Summed and (b) average payoff schemes.
}\label{fig:cgeqd}
\end{center}
\end{figure}

We have set $\gamma_{CC} < \gamma_{DD}$ in the previous numerical simulations
on the basis of the intuition that
neighboring Cs may be more willing to remain connected than
neighboring Ds.
In this case, Cs tend to have larger degrees than Ds.
To show that this assumption is not needed for enhancing
cooperation, we perform numerical simulations by swapping
the values of $\gamma_{CC}$ and $\gamma_{DD}$ .
We use Fermi update rule and the complete graph as the initial network.
The results for the two payoff schemes are shown in
Fig.~\ref{fig:cgeqd}. Although the parameter region for enhanced
cooperation is relatively small, cooperation is viable when AL
dynamics are fast enough and sufficient Cs exist initially.

\subsection{Initially sparse networks}\label{sub:sparse}

\begin{figure}[!t]
\begin{center}
  \begin{tabular}{cc}
\hspace{-2cm}
     \rotatebox{-90}{\resizebox{60mm}{!}{\includegraphics{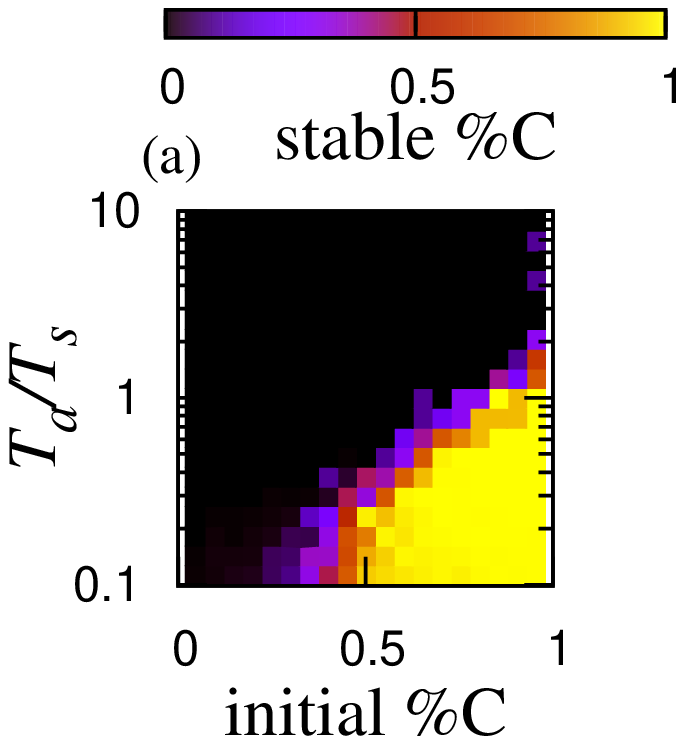}}}
\hspace{-4.5cm}
     \rotatebox{-90}{\resizebox{60mm}{!}{\includegraphics{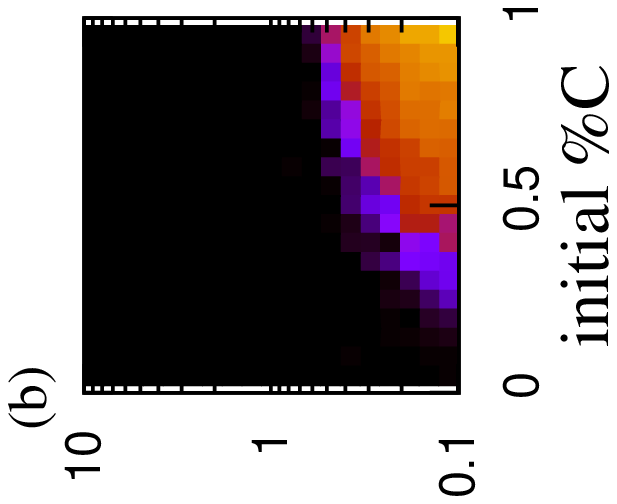}}} \\
  \end{tabular}
\caption{(Color online) Final fraction of cooperators when the initial
network is the random graph with mean degree 8.
We use Fermi update rule.
(a) Summed and (b) average payoff schemes.
}\label{fig:sparse}
\end{center}
\end{figure}

In the previous sections, we performed simulations with the all-to-all
connection initially.
In this section,
we examine the case in which the network is sparse in the beginning.
For Fermi update rule,
the final fraction of cooperators when the initial network
is a random graph with mean degree 8 is shown 
in Fig.~\ref{fig:sparse}. The results are almost the same as 
those shown in Fig.~\ref{fig:al}(a, b).

\section{Conclusions}\label{sec:con}

We have demonstrated that the coevolutionary prisoner's dilemma game promotes
cooperation in the average payoff scheme by extending the 
results for the summed payoff scheme with AL dynamics
\cite{Pacheco_etal_2006a,Pacheco_etal_2006b}.
The results are robust against the introduction of heterogeneous
connectivity
inherent
in the players, changes in 
the update rule, an increase in the rate of removing
C-C links, and  the density of links in the
initial network.

We remark that Fu and colleagues obtained 
the results similar to ours.
Using different link dynamics,
they showed that
cooperation is enhanced 
under both summed and average payoff schemes if
the link dynamics are fast
\cite{Fu_etal2009}. In comparison,
we have shown that the comparable results also hold true for
AL dynamics and that the results are robust with respect to the
heterogeneity in the degree, the update rule, and an increased probability of
pruning C-C links.

\section*{Acknowledgements}
We thank Takehisa Hasegawa for fruitful discussions.
N.M. acknowledges the support through Grants-in-Aid for Scientific Research
(Nos. 20760258, 20540382, and 20115009) from MEXT, Japan.

\bibliography{ybib.bib}
\end{document}